# The Commensurate–Incommensurate Charge-Density-Wave Transition and Phonon Zone Folding in 1$T$-TaSe$_2$ Thin Films

R. Samnakay[1, 2], D. Wickramaratne[3], T. R. Pope[4], R. K. Lake[3], T. T. Salguero[4] and A.A. Balandin[1, 2, #]

[1]Nano-Device Laboratory (NDL), Department of Electrical and Computer Engineering, University of California – Riverside, Riverside, California 92521 USA

[2]Phonon Optimized Engineered Materials (POEM) Center, Materials Science and Engineering Program, University of California – Riverside, Riverside, California 92521 USA

[3]Laboratory for Terascale and Terahertz Electronics (LATTE), Department of Electrical and Computer Engineering, University of California – Riverside, Riverside, California 92521 USA

[4]Department of Chemistry, University of Georgia, Athens, Georgia 30602 USA

R. Samnakay, D. Wickramaratne, T. R. Pope, R. K. Lake, T. T. Salguero and A.A. Balandin (2014)

## Abstract

Bulk 1$T$-TaSe$_2$ exhibits unusually high charge density wave (CDW) transition temperatures of 600 K and 473 K below which the material exists in the incommensurate (I-CDW) and the commensurate (C-CDW) charge-density-wave phases, respectively. The $\sqrt{13} \times \sqrt{13}$ C-CDW reconstruction of the lattice coincides with new Raman peaks resulting from zone-folding of phonon modes from middle regions of the original Brillouin zone back to Γ. The C-CDW transition temperatures as a function of film thickness are determined from the evolution of these new Raman peaks and they are found to decrease from 473K to 413K as the film thicknesses decrease from 150 nm to 35 nm. A comparison of the Raman data with *ab initio* calculations of both the normal and C-CDW phases gives a consistent picture of the zone-folding of the phonon modes following lattice reconstruction. In the I-CDW phase, the loss of translational symmetry coincides with a strong suppression and broadening of the Raman peaks. The observed change in the C-CDW transition temperature is consistent with total energy calculations of bulk and monolayer 1$T$-TaSe$_2$.

#Corresponding author (AAB): balandin@ee.ucr.edu

R. Samnakay, D. Wickramaratne, T. R. Pope, R. K. Lake, T. T. Salguero and A.A. Balandin (2014)

Recent advances in the exfoliation and growth of two-dimensional (2D) layered materials have enabled the investigation of material systems that were previously difficult to study [1-4]. The transition metal dichalcogenides (TMDs) belong to the class of *van der Waals* materials that exhibit a variety of interesting properties including superconductivity [5], thermoelectricity [6-7], low-frequency current fluctuations [8], and charge density waves (CDWs) [9]. The CDW state is a macroscopic quantum state consisting of a periodic modulation of the electronic charge density accompanied by a periodic distortion of the atomic lattice in a quasi-1D or quasi-2D layered metallic crystal [10-15]. In quasi-1D bulk crystals, the transition to the CDW state is conventionally described as the result of Fermi surface nesting giving rise to a CDW with wavevector $Q = 2k_F$ and an energy gap at the Fermi energy. In 2D crystals, the physical mechanism of CDW phase transitions is material dependent [16-18]. In samples of 1*T*-TaS$_2$ and 1*T*-TaSe$_2$, ARPES experiments have shown partial gapping of the Fermi surface but the origin of the transition is still a subject of debate [17]. In 2*H*-NbSe$_2$ the Fermi surface nesting was recently questioned as the origin of the CDW [18] while the origin of the exact mechanism is also a subject of debate. Also, a number of different CDW phases can exist at different temperatures: commensurate CDW (C-CDW), nearly-commensurate CDW (N-CDW) and incommensurate CDW (I-CDW).

The advent of layered van der Waals materials and methods for their exfoliation stimulated interest in quasi-2D CDW materials and thin films derived from these materials. Some of us previously demonstrated that the transition temperature to the CDW phase of TiSe$_2$ can be increased by thinning the bulk crystal into a thin film [9]. The experimental results in Ref. [9] were supported by theoretical considerations [19]. The evolution of the transition temperature depends on the trade-off between the elastic energy required for the deformation of the lattice and the electronic energy reduction. For this reason, the reverse temperature dependence— decreasing transition temperature with decreasing thickness of TMD films— is also possible. A reduction of the C-CDW transition temperature with reducing film thickness has been observed in VSe$_2$ thin films [20] and 1*T*-TaS$_2$ thin films [21-22]. The gate control of the CDW transition has been demonstrated by several groups [21- 22]. The possibility of tuning and controlling CDW phase transitions is motivated by proposed sensor and electronic applications.

R. Samnakay, D. Wickramaratne, T. R. Pope, R. K. Lake, T. T. Salguero and A.A. Balandin (2014)

In this Letter, we experimentally and theoretically study the effect of the C-CDW lattice reconstruction on the Raman spectra and the Γ-point phonon modes. The new Γ-point modes that appear upon lattice re-construction can be viewed as originating from middle regions of the unreconstructed Brillouin zone that are folded back to Γ by integer multiples of the C-CDW wave vector. By measuring the temperature at which the new Raman peaks appear, Raman spectroscopy serves as a metrology tool to measure the temperature dependence of the C-CDW transition as a function of film thickness. Using one particular polymorph of tantalum diselenide, $1T$-TaSe$_2$, we find that the C-CDW transition temperature falls with decreasing film thickness over the range of the film thicknesses studied. As the temperature increases so that the sample transitions from the C-CDW phase to the I-CDW phase, the Raman peaks merge and become very broad consistent with loss of translational symmetry of the lattice. Total energy calculations clarify why the preparation of pure phase $1T$- and $2H$-TaSe$_2$ samples is challenging.

TaSe$_2$ can exist in many different polymorphs and polytypes; the $1T$, $2H$ and $3R$ structures are shown in Figure 1. The two polymorphs with CDW transitions are characterized by trigonal-prismatic ($2H$) and octahedral ($1T$) ordering about the Ta atom. The $1T$ polytype is less common yet quite interesting owing to its $T_C$ and $T_{IC}$, which are substantially above RT. In its normal phase (above $T_{IC}$=600 K), $1T$-TaSe$_2$ is a metal [23-24], with lattice parameters of a ~ 0.3480 nm and c = 0.627 nm [25]. The structure of undistorted $1T$-MX$_2$ compounds corresponds to the $D_{3d}$ space group with the $A_{1g}$ and $E_g$ Raman active modes, which are the focus of the present study. As the temperature T is lowered below $T_{IC}$=600 K, the bulk $1T$-TaSe$_2$ crystal transforms into the I-CDW phase. Decreasing T below $T_C$=473 K results in the first order phase transition to the C-CDW phase. During the I-CDW to C-CDW phase transition, the CDW supercell undergoes a $\sqrt{13} \times \sqrt{13}$ reconstruction in which the basal-plane lattice vectors increase by $\sqrt{13}$ and rotate by 13.9° with respect to the original lattice vectors [26]. The new C-CDW phase has a triclinic structure of the space group $C_i^1$, which has 13 formula units per primitive cell and 114 optical modes with 57 Raman active $A_g$ modes [27]. For this reason, varying temperature over $T_C$ or $T_{IC}$ results in strong modifications of the Raman spectra of $1T$-TaSe$_2$.

[Figure 1: Crystal Structure of 3 Polytypes]

R. Samnakay, D. Wickramaratne, T. R. Pope, R. K. Lake, T. T. Salguero and A.A. Balandin (2014)

For this work, crystals of 1$T$-TaSe$_2$ were grown by the chemical vapor transport (CVT) method [4, 28-31]. Stoichiometric amounts of elemental tantalum and selenium were ground together, sealed in an evacuated fused quartz ampoule with a transport agent (I$_2$), and heated to >850 °C in a two zone tube furnace with one zone set 50–100 °C greater than the other zone. After a suitable growth period, the ampoule was cooled rapidly to trap the metastable 1$T$-TaSe$_2$ structure. This process also resulted in some 3$R$-TaSe$_2$ crystals and residual 2$H$-TaSe$_2$; however, it was possible to differentiate the 1$T$ and 3$R$ crystals by their distinctive gold and silver colors, respectively, and thus to physically separate them (Figure 2). The crystals exhibited excellent crystallinity by powder X-ray diffraction measurements, and EDS and EPMA analyses confirmed the TaSe$_2$ stoichiometry. The 1$T$-TaSe$_2$ thin films were mechanically delaminated from the as-grown 1$T$-TaSe$_2$ crystals. The films were transferred to Si/SiO$_2$ substrates and thinned down using the "graphene-like" mechanical exfoliation approach. Film thicknesses, which ranged from 10 to 400 nm, were measured using atomic force microscopy in the tapping mode regime. Additional details about the preparation and characterization of 1$T$-TaSe$_2$ samples are provided in the *Supplemental Information*.

[Figure 2: Photo of CVT Crystals]

Micro-Raman spectroscopy (Renishaw InVia) was used as a metrology tool in two capacities. First, it was utilized for assessing the material's composition, crystallinity, and thickness. Second, the Raman spectral changes with temperature were recorded to extract the C-CDW phase transition temperature. The variable-temperature Raman spectroscopy has been used as a method of choice for detecting the phase transition temperature in CDW materials, superconductors and other materials undergoing structural phase transitions [32-37]. The Raman measurements were performed in the backscattering configuration under $\lambda$=633 nm laser excitation. We avoided Raman measurements on individual tri-layers of 1$T$-TaSe$_2$ because of laser heating at even small excitation levels for such samples. The bulk crystals of 1$T$-TaSe$_2$ have a low thermal conductivity of ~ 8 W/mK at RT [4], and the thermal conductivity of the thin films is decreased further due to the acoustic phonon scattering from top and bottom interfaces [38-39]; unlike graphene, other exfoliated van der Waals materials do not have atomically smooth interfaces without surface roughness. To avoid local heating, the excitation power for Raman spectroscopy measurements was kept below 0.5 mW. To ensure that no oxidation occurred at elevated temperatures, all measurements were performed with the samples under an argon atmosphere. The temperature was varied from 213 K to 493 K, and Raman spectra were recorded during the heating and cooling cycles within this range. Upon cooling, the Raman features observed at RT were restored. This ensured that the Raman spectral modifications were due to changes in the material's crystal structure (C-CDW to I-CDW phase transition) rather than oxidation.

The polytype and crystallinity of 1$T$-TaSe$_2$ samples were verified by the presence of two characteristic peaks located at ~177 cm$^{-1}$ and ~187 cm$^{-1}$ in agreement with literature values for bulk 1$T$-TaSe$_2$ [26, 40-41]. The ~187 cm$^{-1}$ peak is assigned as A$_{1g}$ [26] whereas the peak at ~177 cm$^{-1}$ is most likely of E$_{2g}$ symmetry (Figure 3 (a)). In comparison, the 2$H$ polytype of TaSe$_2$ has peaks at 207 cm$^{-1}$ (E$_{2g}$) and 235 cm$^{-1}$ (A$_{1g}$) [4], which are substantially different and provide reliable identification of 1$T$- versus 2$H$-TaSe$_2$ using Raman spectroscopy. The ratio of the intensity of the Si substrate peak to that of A$_{1g}$ peak was used to determine the thickness of the exfoliated films. Figure 3 (b) shows the ratio of the intensity of the Si peak at 522 cm$^{-1}$ to that of the A$_{1g}$ peak, I(Si)/I(A$_{1g}$), as a function of the thickness H. The I(Si)/I(A$_{1g}$) curve was calibrated with AFM thickness measurements and then used for *in situ* extraction of thickness data. This ratio decreases exponentially with increasing film thickness and can be fitted with the equation I(Si)/I(A$_{1g}$)=553exp{-H[nm]/2.9}+4. The decrease in the Si peak intensity is explained by the reduction of the interaction volume inside the Si wafer with increasing thickness of the film on top. This technique has been used for many van der Waals materials [42-44]. The presence of both 1$T$- and 2$H$-TaSe$_2$ in our samples (along with a small amount of the 3$R$ polytype) facilitated the temperature dependent Raman studies: the 2$H$ peaks fortuitously provided a convenient internal reference because its C-CDW transition temperature is substantially below RT (T$_C$ = 90 K).

[Figure 3: Raman Characterization]

To investigate the evolution of the Raman spectrum of 1$T$-TaSe$_2$ thin films near T$_C$, we placed the samples in a hot-cold cell under an argon atmosphere. The sample temperature was varied from 213 K to 493 K, and unpolarized Raman spectra were collected at ~20 different

R. Samnakay, D. Wickramaratne, T. R. Pope, R. K. Lake, T. T. Salguero and A.A. Balandin (2014)

temperatures for each sample. The temperature then was reduced from 493 K to 303 K and Raman spectra were taken again to ensure that no oxidation occurred at higher temperatures. Figure 4 shows representative Raman spectra from a 1$T$-TaSe$_2$ film with a thickness H ~ 150 nm during the heating cycle from 213 K to 493 K. Due to its relatively large thickness, this sample can be considered a "bulk" reference. The two distinctive 1$T$-TaSe$_2$ peaks are seen clearly at ~175 - 178 cm$^{-1}$ and ~185 - 188 cm$^{-1}$ [26-27, 40-41]. The Raman peaks at 209 - 212 cm$^{-1}$ and 233 - 236 cm$^{-1}$ correspond to the E$_{2g}$ and A$_{1g}$ modes of 2$H$-TaSe$_2$. The origin of the pronounced peak at ~154 cm$^{-1}$ has not been described previously in the literature.

[Figure 4: Raman of "Bulk" in Heating Cycle]

One can see from Figure 4 the strong changes in the Raman spectra at around T$_C$= 473 K for this "bulk" sample. These changes were reproducible for several experimental runs and samples. As the temperature approaches T$_C$=473 K, the two characteristic 1$T$-TaSe$_2$ peaks at ~175 - 178 cm$^{-1}$ and ~185 - 188 cm$^{-1}$ merge together, forming one broad peak (e.g., the blue curve in Figure 4). The second modification as the material enters the I-CDW phase is a strong broadening of the Raman peaks and reduction in their intensity. The spectra of 1$T$-TaSe$_2$ in the I-CDW phase (above T$_C$=473 K) appear more as continua than sets of narrow peaks. This feature can be explained by symmetry considerations. In the I-CDW phase, the translation symmetry of the lattice is lost and the components of the phonon momentum parallel to the I-CDW distortions are no longer good quantum numbers. For this reason, the momentum conservation rules for the first-order Raman scattering to Brillouin zone (BZ) center phonons are relaxed, and as a result, all phonons can take part in the scattering. The loss of translational symmetry and corresponding relaxation of the phonon momentum conservation in the I-CDW phase is somewhat similar to that in amorphous materials, which also are characterized by broad, continuum-type Raman spectra. The unknown peak at 154 cm$^{-1}$ undergoes a similar evolution (intensity decrease with corresponding broadening) as the E$_{2g}$ and A$_{1g}$ symmetry peaks of 1$T$-TaSe$_2$. Although the 2$H$ peaks undergo some broadening with increasing temperature, they are not affected as much and remain as two separate peaks throughout the temperature range examined here.

R. Samnakay, D. Wickramaratne, T. R. Pope, R. K. Lake, T. T. Salguero and A.A. Balandin (2014)

Figure 5 shows Raman spectra collected during the cooling cycle of the same "bulk" $1T$-TaSe$_2$ sample. As the temperature was decreased from 493 K to 303 K, at $T_C$=473 K the combined peak splits into the two constituent $1T$ polytype peaks at ~175 - 178 cm$^{-1}$ and ~185 - 188 cm$^{-1}$, and the intensity of these peaks increases, approaching the RT values. The unknown peak at ~154 cm$^{-1}$ also reappears at the same temperature during the cooling cycle. The RT Raman spectrum of $1T$-TaSe$_2$ again becomes a set of relatively narrow peaks as expected for materials with translational symmetry in the C-CDW state. The restoration of the original spectrum of the sample after heating and cooling cycles proves that the spectral changes are reversible and not a result of oxidation or surface contamination.

[Figure 5: Raman of "Bulk" in Cooling Cycle]

Figure 6 shows Raman spectra from a much thinner flake with H ~ 35 nm. One can see the same characteristic peaks of $1T$-TaSe$_2$ at ~ 177 cm$^{-1}$ and ~189 cm$^{-1}$ as well as the two $2H$-TaSe$_2$ peaks at ~210 cm$^{-1}$ and ~231 cm$^{-1}$. As the sample temperature increases, the two $1T$-TaSe$_2$ Raman peaks begin to merge, similar to what was observed for the "bulk" sample. However, the merging rate clearly is enhanced, and the temperature at which the two peaks completely merge is reduced to $T_C$=413 K. Thus the spectral modification temperature has been shifted by about 13%. These data indicate that the temperature of the transition between the C-CDW and I-CDW phases decreases with decreasing thickness of the $1T$-TaSe$_2$ film. Figure 7 shows the cooling cycle for the same sample (H=35 nm) to prove that no oxidation or other changes occurred during the heating cycle, which could impact the Raman spectra. At RT, the original Raman spectrum of the sample has been restored completely. The presence of the $2H$ polytype in our sample does not affect the conclusion because $2H$-TaSe$_2$ has much lower transition temperatures of $T_C$=90 K and $T_{IC}$=122 K compared to $1T$-TaSe$_2$. Table I summarizes the dependence of $T_C$ on $1T$-TaSe$_2$ sample thickness.

[Figure 6: Raman of the Thin Film in Heating Cycle]

[Figure 7: Raman of the Thin Film in Cooling Cycle]

R. Samnakay, D. Wickramaratne, T. R. Pope, R. K. Lake, T. T. Salguero and A.A. Balandin (2014)

**Table I:** Experimental Transition Temperature to C-CDW Phase

| Sample Thickness H (nm) | Transition Temperature $T_C$ (K) |
|---|---|
| 150 | 468-473 |
| 55 | 413-433 |
| 41 | 403-423 |
| 35 | 393-413 |
| 18 | ~353 |

To further rationalize the experimental data in quasi-2D CDW crystals, we calculated the total energy of bulk and monolayer 2H- and 1T-TaSe$_2$ in their normal (existing for T>T$_{IC}$) and C-CDW (existing for T<T$_C$) phases. Our calculations were based on the first-principles density functional theory (DFT) using the projector augmented wave method as implemented in the software package VASP [45] and density functional perturbation theory (DFPT) as implemented in the Quantum-ESPRESSO package. The details and parameters of our calculations are described in the *METHODS* section and relevant literature [25, 46-47].

We first compare the relative stability of the 2H and 1T polytypes of bulk TaSe$_2$. The 2H polytype is the ground state stacking order and is lower in energy by 0.121 eV/Ta-atom compared to the 1T polytype. This energy difference is a factor of 7 lower than the energy barrier between the 2H and 1T polytypes of MoS$_2$ [48]. This small difference explains why the material grown by CVT contains 2H-TaSe$_2$ even when the synthesis process is optimized for 1T growth (i.e., rapid quenching). The differences in the ground state energies of the C-CDW phases and the normal phases of bulk and monolayer 1T and 2H TaSe$_2$ also were calculated. Figure 8 illustrates the C-CDW structure for each polytype: 2H-TaSe$_2$ undergoes a commensurate ($3a_O \times 3a_O \times 1$) periodic lattice distortion while 1T-TaSe$_2$ undergoes a commensurate ($\sqrt{13}a_O \times \sqrt{13}a_O \times 1$) periodic lattice distortion. For each calculation of the bulk and monolayer C-CDW reconstructed lattices, the lattice constants were fixed and the atomic coordinates were allowed to relax after the tantalum atoms had been displaced from their equilibrium positions.

[Figure 8: Schematic of the CDW Lattice Structure]

The results of these total energy calculations are summarized in Table II, where $\Delta E = E_N - E_{C-CDW}$ is the difference between the ground state energies of the normal structure and the C-CDW structure. For each polytype and dimension, the commensurate CDW structure is predicted by DFT to be the ground state structure. The energy reduction of the bulk CDW supercell is greater than the energy reduction of the monolayer CDW supercell for both the 1$T$ and 2$H$ polytypes. These results are consistent with prior reports of total energy calculations on the bulk C-CDW supercells in 2$H$- and 1$T$-TaSe$_2$ [49, 24]. A transition to a commensurate CDW lowers the electronic energy [10]. A smaller energy reduction |ΔE| indicates a lower transition temperature. The results of the total energy calculations on 2$H$-TaSe$_2$ are consistent with prior *ab-initio* calculations of the energy reduction in the bulk and monolayer C-CDW supercells of 2$H$-TaSe$_2$ [49].

**Table II:** Calculated Energy Reduction in 1$T$ and 2$H$ TaSe$_2$ C-CDW

| ΔE (meV/Ta-atom) | Bulk  | 1L    |
| ---------------- | ----- | ----- |
| 1$T$-TaSe$_2$    | -21.9 | -19.4 |
| 2$H$-TaSe$_2$    | -3.50 | -2.04 |

In order to explain the peak that occurs at 152-154 cm$^{-1}$, the single q-point phonon frequencies of bulk and monolayer TaSe$_2$ were calculated using density functional perturbation theory (DFPT). The reconstruction of the lattice along the basal plane in the C-CDW state of TaSe$_2$ results in a reduced Brillouin zone of the C-CDW structure (C-BZ). The C-BZ forms a subset of the normal Brillouin zone (N-BZ) of TaSe$_2$ in its normal undistorted state. Twelve points in the N-BZ get mapped back to Γ in the C-BZ. Figure 9 shows the structure of the N-BZ and the C-BZ and the 12 q-points in the N-BZ that are zone-folded to the Γ point of the C-BZ. These zone-folded modes can result in new peaks in the Raman spectrum. The 12 q-points in the N-BZ that map back to Γ in the C-BZ are ±$\mathbf{g}_1$, ±$\mathbf{g}_2$, ±($\mathbf{g}_1$ - $\mathbf{g}_2$), ±($\mathbf{g}_1$ + $\mathbf{g}_2$), ±($\mathbf{g}_2$ - 2$\mathbf{g}_1$) and ±(2$\mathbf{g}_2$ - $\mathbf{g}_1$) where $g_1$ and $g_2$ are the reciprocal lattice constants of the C-BZ. The reciprocal lattice constants are $\mathbf{g}_1 = \frac{G}{13}\left(2\sqrt{3},\ -1\right)$ and $\mathbf{g}_2 = \frac{G}{13}\left(-\sqrt{3}/2,\ 7/2\right)$, respectively, where G is the magnitude of the reciprocal lattice vector of the N-BZ. The normal-phase phonon frequencies at Γ, $\mathbf{q}$ = $\mathbf{g}_1$ and $\mathbf{q}$ = $\mathbf{g}_1$+$\mathbf{g}_2$ in the N-BZ are calculated and summarized in Table III for bulk and monolayer 1$T$-TaSe$_2$.

R. Samnakay, D. Wickramaratne, T. R. Pope, R. K. Lake, T. T. Salguero and A.A. Balandin (2014)

The 5 other q-points shown in Figure 9 with magnitude $|g_1|$ have the same frequencies as those at $g_1$, and the 5 other q-points with magnitude $|g_1+g_2|$ have the same frequencies as those at $g_1+g_2$. Only frequencies between 100 cm$^{-1}$ and 270 cm$^{-1}$ are shown.

**Table III:** Calculated phonon frequencies of normal-phase bulk and monolayer 1$T$-TaSe$_2$ at Γ and at the q points in the N-BZ that are folded back to Γ in the C-BZ. The IR active (IR) and Raman active (**R**) modes at Γ are indicated.

| q-point | Monolayer normal phase (cm$^{-1}$) | Bulk normal phase (cm$^{-1}$) |
|---|---|---|
| Γ | 143 (IR), 143 (IR), 173 (**R**), 173(**R**), 188 (**R**), 210 (IR) | 146 (IR), 146 (IR), 175 (**R**), 175 (**R**), 191(**R**), 219 (IR) |
| $g_1$ | 109, 118, 144, **154**, 180, 249 | 122, 149, **152**, 197, 215, 261 |
| $g_1+g_2$ | 108, 117, 143, **154**, 165, 181, 249 | 121, 148, **153**, 172, 196, 214, 260 |

As shown in Table III, the modes originating from the q-points in Figure 9 have energies of ~152 – 154 cm$^{-1}$ (highlighted in bold font). This indicates that all of these phonon modes may contribute to the Raman peak observed experimentally at 152 – 154 cm$^{-1}$. The fact that the 12 q-points in the N-BZ that are equivalent to Γ in the C-BZ have frequencies 152 – 154 cm$^{-1}$ corresponding to the new Raman peak is consistent with the view that the new peak is the result of zone-folding associated with the re-constructed lattice. We also observe that the the Raman active modes of the normal bulk phase are close to the experimentally observed modes at 177 cm$^{-1}$ and 187 – 189 cm$^{-1}$ shown in Figs. 4-7. So far, we have considered phonon modes of the normal lattice consisting of 3 atoms per unit cell at momenta that are at equivalent Γ points of the reconstructed C-CDW lattice, and we find that one of their energies corresponds to a new peak in the Raman spectrum of the C-CDW phase.

To further investigate the new Raman peaks of the C-CDW phase, we calculate the phonon energies of the C-CDW reconstructed lattice with 39 atoms in the unit cell for bulk and monolayer 1$T$-TaSe2. Table IV shows the calculated phonon frequencies between 100 cm$^{-1}$ and

R. Samnakay, D. Wickramaratne, T. R. Pope, R. K. Lake, T. T. Salguero and A.A. Balandin (2014)

270 cm$^{-1}$ which is a subset of the 117 phonon modes that occur in the bulk and monolayer C-CDW structures. The modes are grouped according to the phonon energies of the normal phase listed in Table III. The Γ-point phonon energies of the C-CDW phase are centered around the normal-phase phonon energies listed in Table III for both the bulk and monolayer structures. The number of phonon modes that occur within each frequency range is listed in parentheses. The splitting of the phonon energies is not unexpected since the 12 modes of the normal lattice are coupled by the C-CDW wavevectors **Q** which form the reciprocal lattice vectors of the commensurate Brillouin zone (C-BZ) shown in Fig. 9. Thus, the CDW potential V$_Q$ couples and splits the phonon modes of the normal lattice. The calculated bulk Γ-valley phonons listed in Table IV are within 1 cm$^{-1}$ to 6 cm$^{-1}$ of the zone folded normal-phase 152 cm$^{-1}$ frequency given in Table III and the experimentally observed 154 cm$^{-1}$ frequency shown in the Raman spectra in Figs. 4-7. In the bulk and monolayer C-CDW structures, the eigenvectors of the phonon frequencies in the range of 150 cm$^{-1}$ to 158 cm$^{-1}$ are a mixture of transverse and longitudinal displacements of the 26 Se atoms and 13 Ta atoms. As a result, they cannot be easily categorized as A$_{1g}$ or E$_{2g}$ modes, and their symmetry is not obvious. Assuming that these modes originate from a zone-folding of the low-symmetry points **g**$_1$ and **g**$_1$+**g**$_2$ in the N-BZ, this mix of displacements might be expected. The eigenvectors of the normal-phase phonons at **g**$_1$ and **g**$_1$+**g**$_2$ are also a mix of transverse and longitudinal displacements,

**Table IV:** Calculated Γ point phonons of the reconstructed C-CDW structure in monolayer and bulk 1*T*-TaSe$_2$ grouped by the normal-phase phonon energies in Table III.

| C-CDW Γ-point phonons (cm$^{-1}$) ||
| --- | --- |
| **Monolayer 1*T*-TaSe$_2$ C-CDW** | **Bulk 1*T*-TaSe$_2$ C-CDW** |
| 101 - 122 (15) | 120 - 123 (9) |
| 144 (9) | **151 - 158 (15)** |
| **150 - 158 (12)** | 163 - 169 (9) |
| 160 - 179 (13) | 172 - 179 (8) |
| 181 - 189 (11) | 180 - 188 (9) |
| 243 - 257 (18) | 190 - 197 (7) |
| | 201 - 217 (11) |
| | 251 - 266 (10) |

The presence of a zone-folded Raman peak due to a C-CDW lattice distortion also was predicted in bulk 1$T$-TiSe$_2$ [50] and experimentally found in bulk 1$T$-TiSe$_2$ [51]. Reconstruction of the BZ in the C-CDW phase and the appearance of the zone-folded Raman peaks also are reminiscent of the zone folding in twisted bilayer graphene [52-54]. The zone-folded phonon peaks in the C-CDW materials appear at much higher wave numbers than the peaks in man-made quantum well superlattices (QWS) owing to the smaller period of the lattice distortion in a C-CDW. The folded peaks in molecular-beam-epitaxy grown QWS have been typically observed at frequencies below 80 cm$^{-1}$ [39, 55]. In this sense quasi-2D CDW materials present a unique opportunity to investigate zone-folded phonons with conventional Raman spectrometers that have cut-off frequencies of around 100 cm$^{-1}$. In addition the possibility of tuning the incommensurate to commensurate CDW transition temperature can have important implications for the proposed applications of CDW materials in electronics [3, 56].

Since there is a possibility of 3$R$-TaSe$_2$ residue in CVT synthesized samples we theoretically analyzed the phonon spectrum of this polytype to exclude the possibility of the new Raman peaks arising from the 3$R$-TaSe$_2$ phonon modes. We calculated the Γ-point phonons and Raman active modes of bulk 3$R$-TaSe2 using density functional perturbation theory. The 3$R$-TaSe$_2$ polytype has a space group R$_{3m.}$ The unit cell consists of three monolayers of TaSe$_2$ that have a trigonal prismatic coordination of the Ta atom. The Raman modes of the 3$R$ structure remain the out-of-plane A$_{1g}$ mode and the in-plane E$_{2g}$ mode. The calculations indicate that the A$_{1g}$ and E$_{2g}$ frequencies of bulk 3$R$-TaSe$_2$ are ~230 cm$^{-1}$ and ~191 cm$^{-1}$, respectively. These frequencies are not in the range of the new Raman peak of interest and hence do not affect the interpretation of the zone-folded peak.

It is interesting to note that the thickness required for inducing changes to the transition temperature is comparable to the phonon mean free path (MFP). This is the same length scale

R. Samnakay, D. Wickramaratne, T. R. Pope, R. K. Lake, T. T. Salguero and A.A. Balandin (2014)when one starts observing modifications in the phonon spectrum of thin films as compared to bulk dispersion [39]. The trend in the thickness dependence observed for one TMD material cannot be readily extended to other materials from the same group, e.g. $1T$-TaS$_2$. The $1T$-TaSe$_2$ and $1T$-TaS$_2$ polytypes have two different CDW phase diagrams. In bulk $1T$-TaS$_2$, the normal to incommensurate transition occurs at 550 K, an incommensurate to nearly commensurate transition occurs at 350 K and the nearly commensurate to commensurate transition occurs at 180K. A recent study also found a 'super-cooled' state in nanometer thick crystals of $1T$-TaS$_2$ resulting from rapid cooling of their samples, which indicates a suppression of the commensurate CDW state [21].

In summary, during cooling of bulk $1T$-TiSe2 samples, a new Raman peak appeared at 154 cm$^{-1}$ at the known bulk C-CDW transition temperature. The appearance of this peak during cooling then served as an indicator of the C-CDW phase transition. The peak was robust and reversible with alternating temperature sweeps. The temperature at which this peak occurred decreased with decreasing sample thickness indicating a decrease in C-CDW transition temperature with sample thickness. Specifically, the C-CDW transition temperature was determined to decrease from 473 K to 413 K as the film thicknesses decreased from 150 nm to 35 nm. A comparison of the Raman data with *ab initio* calculations of the vibrational modes of both the normal and C-CDW phases gave a consistent picture of the zone-folding of the phonon modes following lattice reconstruction. The q-points of the normal lattice that lie at equivalent Γ points of the C-CDW reconstructed lattice all have a phonon mode at, or within 2 cm$^{-1}$, of the frequency of the new Raman peak. The calculated Γ-point phonon frequencies of the bulk reconstructed C-CDW lattice show a splitting of the modes into a cluster of frequencies between 151 and 158 cm$^{-1}$. These modes are a mixture of transverse and longitudinal displacements of the 26 Se atoms and 13 Ta atoms, and, as a result, they cannot be easily categorized as $A_{1g}$ or $E_{2g}$ modes, and their symmetry is not obvious. The Raman spectra shows a sudden broadening and loss of features as the sample crosses from the C-CDW phase to the I-CDW phase. In the incommensurate phase, the CDW wavevector is incommensurate with the lattice wavevectors, so that the crystal loses its periodicity and translational symmetry. The loss of translational symmetry in the I-CDW phase coincides with a strong suppression and broadening of all $1T$-phase Raman peaks. The total energy calculations of the normal and C-CDW phases of bulk and monolayer $1T$ and $2H$

14 | P a g e

polytypes show that the energy difference between the normal and the C-CDW phase of the bulk is greater than that of the monolayer, and this is consistent with a reduced transition temperature for a monolayer compared to that of the bulk. The experimental thicknesses did not come close to a monolayer, so we simply note that the trend of lower C-CDW temperature for lower thickness is consistent with the experimental trend. Overall, this work clarifies and provides insight into the effect of the I-CDW and the C-CDW lattice reconstruction on the Raman spectra, and it serves as a guide for the use of Raman spectroscopy as a metrology tool for characterizing CDW transitions.

## METHODS

The calculations included DFT using the projector augmented wave method as implemented in the software package VASP [45] and DFPT as implemented in the Quantum-ESPRESSO package. For the electronic structure calculations, a Monkhorst-Pack scheme was adopted to integrate over the BZ with a k-mesh 12 × 12 × 1 (12 × 12 × 6) for the monolayer (bulk) structures. A plane-wave basis kinetic energy cutoff of 500 eV was used. The van der Waals interactions in $TaSe_2$ were accounted for using a semi-empirical correction to the Kohn-Sham energies when optimizing the bulk structure [46]. The optimized lattice parameters for bulk $2H$- and $1T$-$TaSe_2$ are a=3.45 Å, c=13.06 Å and a=3.42 Å, c=6.22 Å, respectively. These structural parameters are consistent with prior experimental [25, 47] reports of the lattice parameters for the $2H$ and $1T$ structures. The lattice constants for the monolayer $2H$- and $1T$-$TaSe_2$ structures were obtained from the respective optimized bulk structures. The atomic coordinates within the monolayer $TaSe_2$ structures were optimized by introducing a 20-Å vacuum layer between the adjacent structures. Spin-orbit coupling was included self-consistently in each calculation. For the phonon dispersion calculations using Quantum-ESPRESSO, an energy cutoff of 500 eV was used in the plane wave basis for the bulk and monolayer structures,. The structures were optimized until the forces on the atoms were less than 0.005 eV/Å. A 12×12×6 (12×12×1) k-point grid and a Gaussian smearing of 0.05 eV was used to integrate over the electronic states in the Brillouin zone for the bulk (monolayer) structures. The dynamical matrices for the bulk (monolayer) structure were calculated using a 4×4×2 (4×4×1) q-point grid.

*Acknowledgements*

Theoretical analysis and material characterization were supported by the National Science Foundation (NSF) and SRC Nanoelectronic Research Initiative (NRI) for the project 2204.001: Charge-Density-Wave Computational Fabric: New State Variables and Alternative Material Implementation (NSF ECCS-1124733) as a part of the Nanoelectronics for 2020 and beyond (NEB-2020) program. Material synthesis was supported by the Emerging Frontiers of Research Initiative (EFRI) 2-DARE project: Novel Switching Phenomena in Atomic $MX_2$ Heterostructures for Multifunctional Applications (NSF 005400). Ab initio calculations used the Extreme Science and Engineering Discovery Environment (XSEDE), which is supported by the NSF grant OCI-1053575.

*Authors Contributions*

A.A.B. coordinated the project and led the experimental data analysis and the writing of the manuscript; R.K.L. led the theoretical analysis and contributed to writing of the manuscript; T.T.S. supervised materials synthesis, analyzed experimental data and contributed to writing of the manuscript; R.S. performed exfoliation and temperature-dependent Raman measurements; D.W. conducted *ab initio* simulations and contributed to writing of the manuscript; T.R.P. synthesized material and conducted material characterization.

R. Samnakay, D. Wickramaratne, T. R. Pope, R. K. Lake, T. T. Salguero and A.A. Balandin (2014)

**FIGURE CAPTIONS**

**Figure 1:** Crystal structures of three TaS$_2$ polytypes: 1$T$-TaSe$_2$, 2$H$-TaSe$_2$, and 3$R$-TaSe$_2$. These views show the structures along the a-axis (at left) and c-axis (at right).

**Figure 2:** As-grown crystals of TaSe$_2$: the silver crystals (right) are mixed 2$H$,3$R$-TaSe$_2$ and the golden crystals (left) are 1$T$-TaSe$_2$.

**Figure 3:** (a) Raman spectrum of 1$T$-TaSe$_2$ showing the two identifying peaks at ~177 cm$^{-1}$ and ~187 cm$^{-1}$. (b) The ratio of the intensity of Si peak to A$_{1g}$ peak in Raman spectrum of 1$T$-TaSe$_2$. The calibrated intensity ratio was used for nanometrology of TaSe$_2$ films. The data are from films ranging from 10 nm to 400 nm in thickness. The inset shows A$_{1g}$ and E$_g$ vibrational modes.

**Figure 4:** Temperature-dependent Raman spectrum of the "bulk" (150 nm thick) 1$T$-TaSe$_2$ sample in the heating cycle from 213 K to 493 K. The main spectral features are indicated in the legends and in the text. Note that above T$_{IC}$=473 the spectrum is strongly modified owing to the loss of translational symmetry.

**Figure 5:** Temperature-dependent Raman spectrum of the same "bulk" (150 nm thick) sample in the cooling cycle from 493 K to 303 K. The main spectral features are indicated in the legends and in the text. A complete restoration of the room-temperature Raman spectrum confirms that no oxidation or surface contamination had occurred at high temperatures.

**Figure 6:** Temperature-dependent Raman spectrum of the thin (35 nm thick) 1$T$-TaSe$_2$ sample in the heating cycle from 213 K to 493 K. The main spectral features are indicated in the legends and in the text. Note that the transition temperature between the incommensurate and commensurate phases has lowered to T$_{IC}$=413. The spectrum is strongly modified in the IC-CDW phase owing to the loss of translational symmetry.

R. Samnakay, D. Wickramaratne, T. R. Pope, R. K. Lake, T. T. Salguero and A.A. Balandin (2014)

**Figure 7:** Temperature-dependent Raman spectrum of the same thin film (35 nm thick) sample in the cooling cycle from 493 K to 303 K. The main spectral features are indicated in the legends and in the text. A complete restoration of the room-temperature Raman spectrum confirms that no oxidation or surface contamination had occurred at high temperatures.

**Figure 8:** Commensurate CDW structures of $TaSe_2$: (a) $\sqrt{13}\times\sqrt{13}$ structure of $1T$-$TaSe_2$ and (b) $3\times3$ structure of $2H$-$TaSe_2$.

**Figure 9:** Normal (red) and commensurate reconstructed (green) Brillouin zone of bulk and monolayer $TaSe_2$. The equivalent $\Gamma$ points in the first extended C-BZ and second extended C-BZ are connected to $\Gamma$ by red and blue vectors, respectively.

R. Samnakay, D. Wickramaratne, T. R. Pope, R. K. Lake, T. T. Salguero and A.A. Balandin (2014)

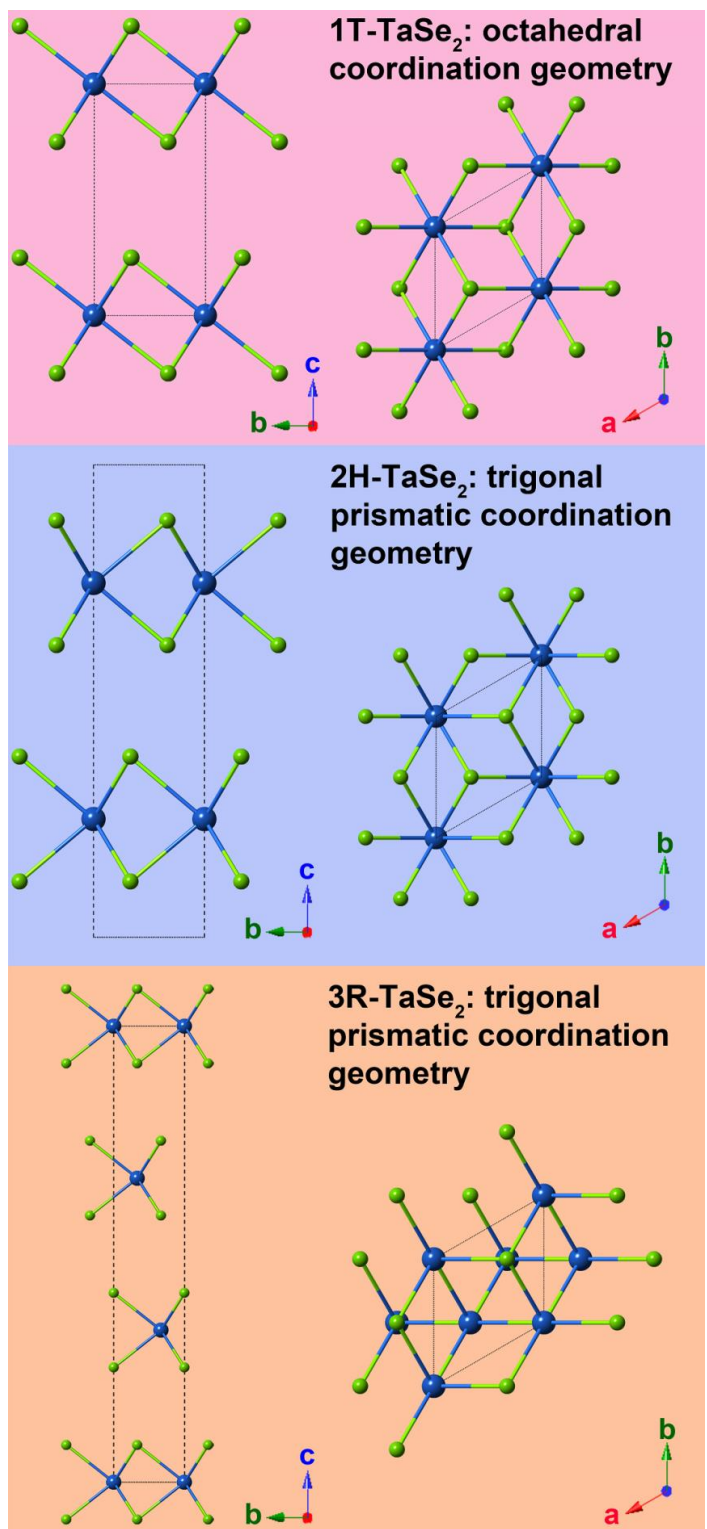

Figure 1

R. Samnakay, D. Wickramaratne, T. R. Pope, R. K. Lake, T. T. Salguero and A.A. Balandin (2014)

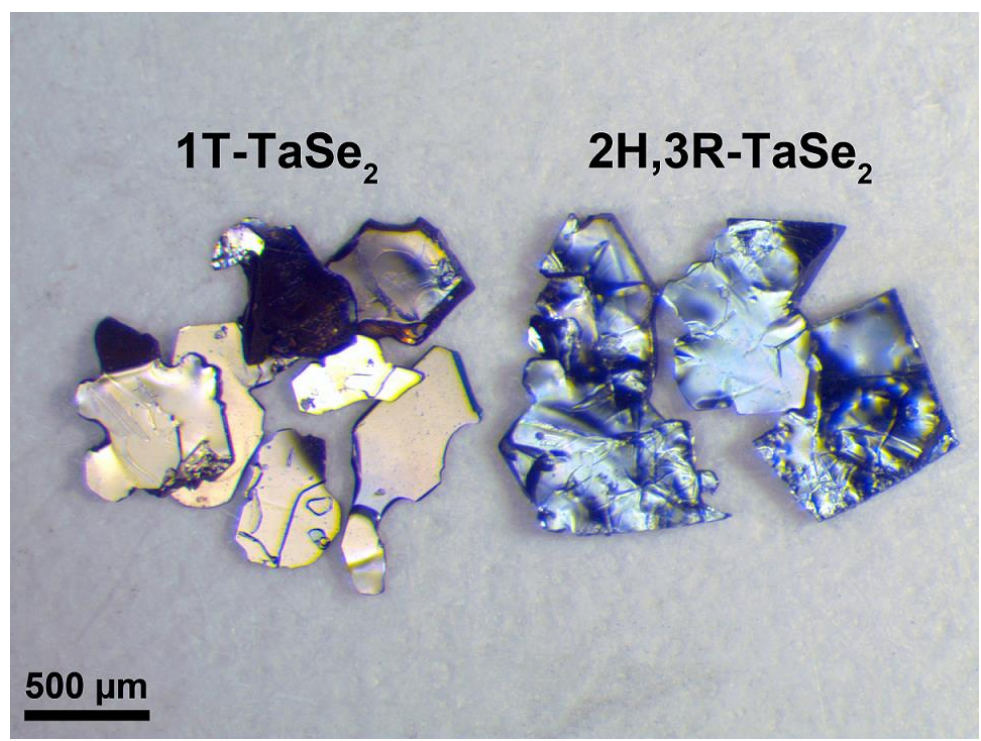

Figure 2

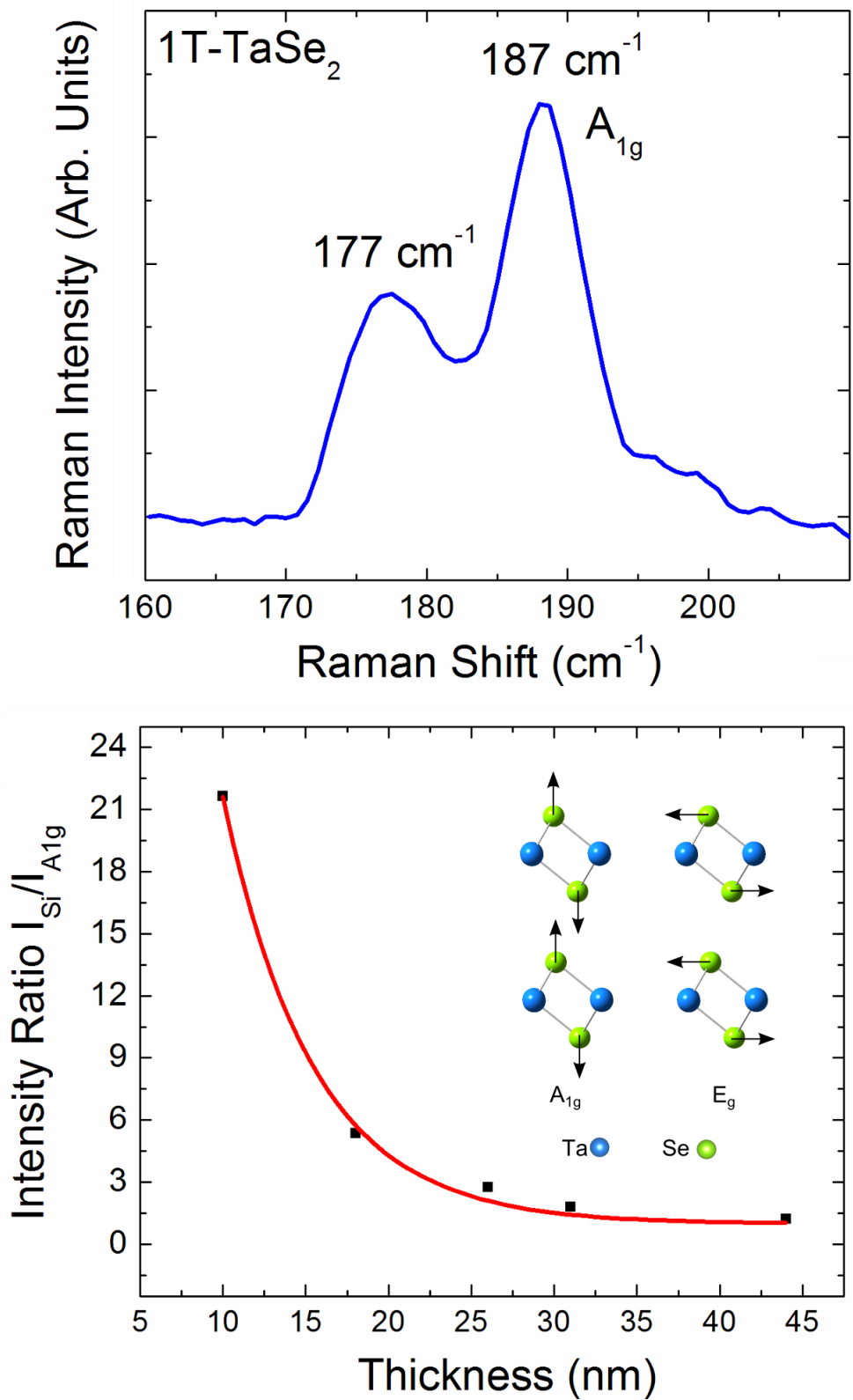

Figure 3

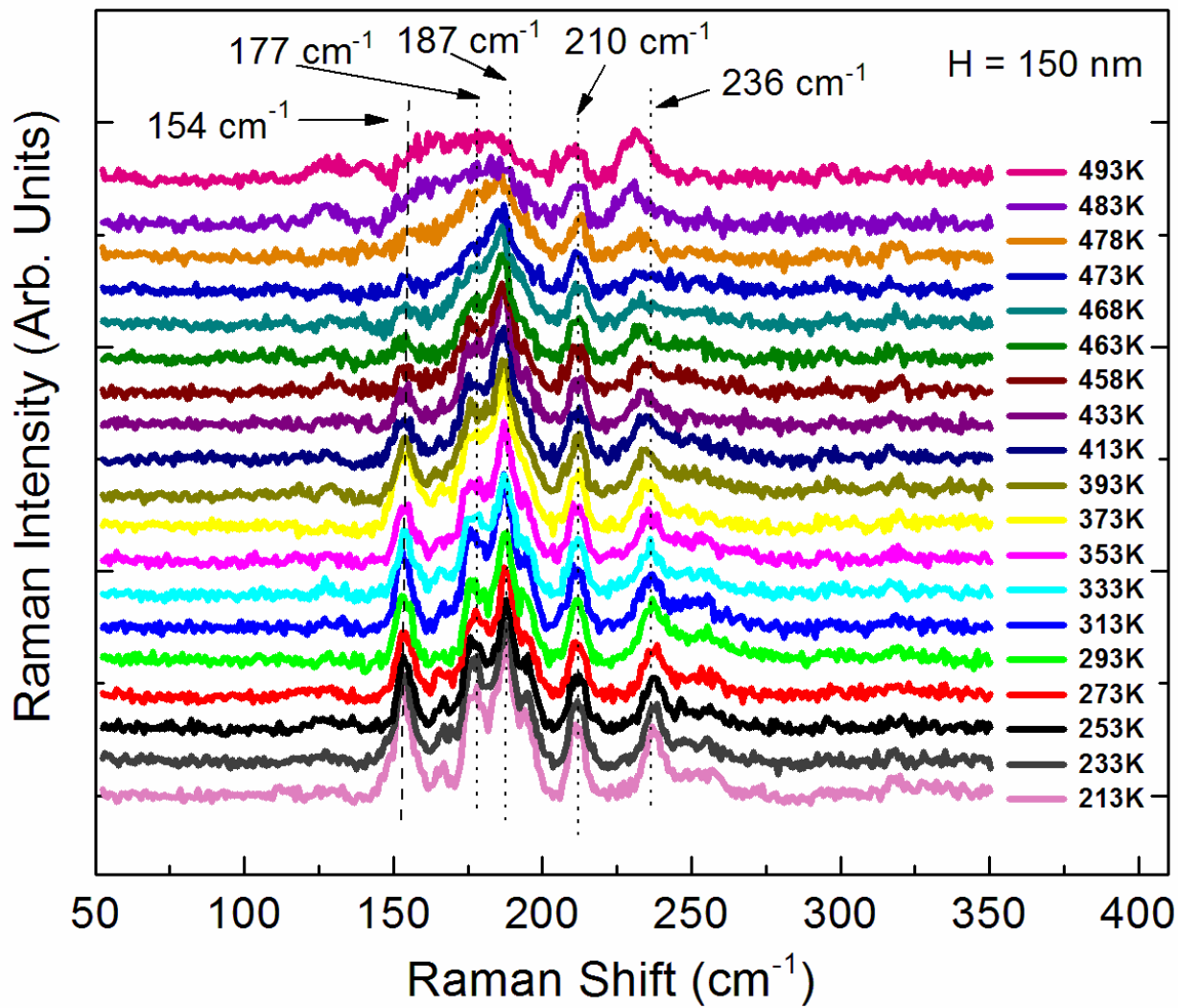

Figure 4

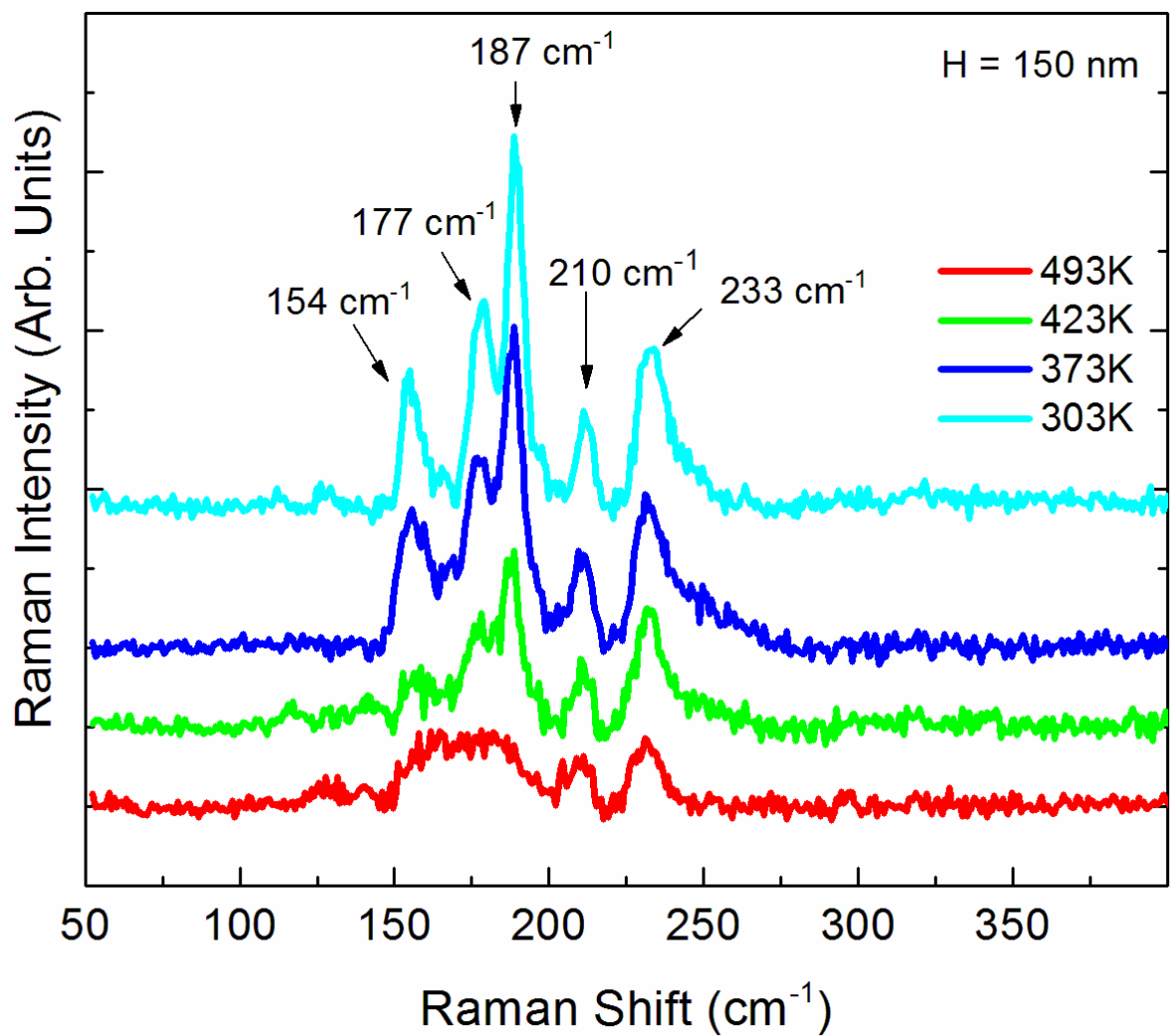

Figure 5

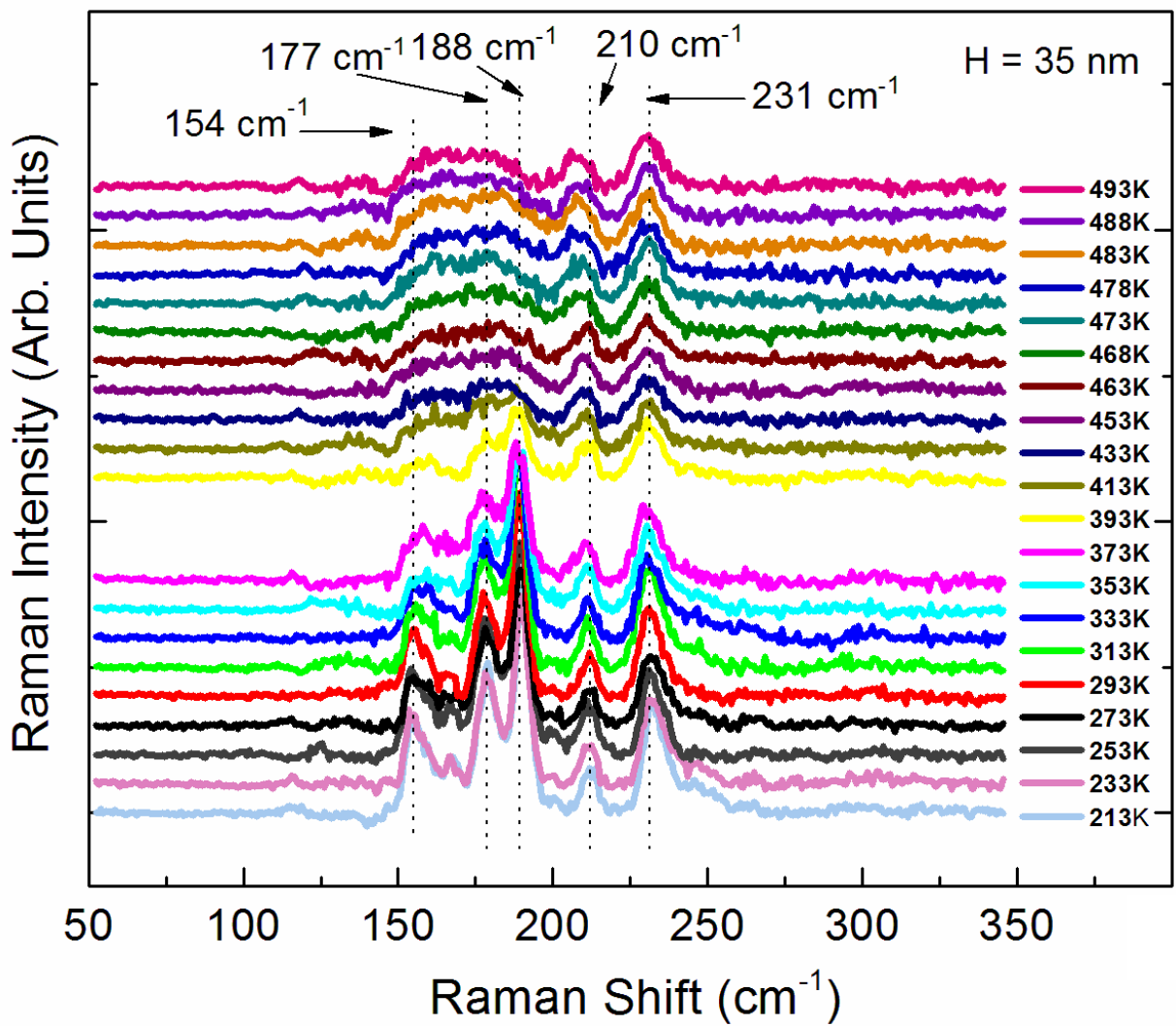

Figure 6

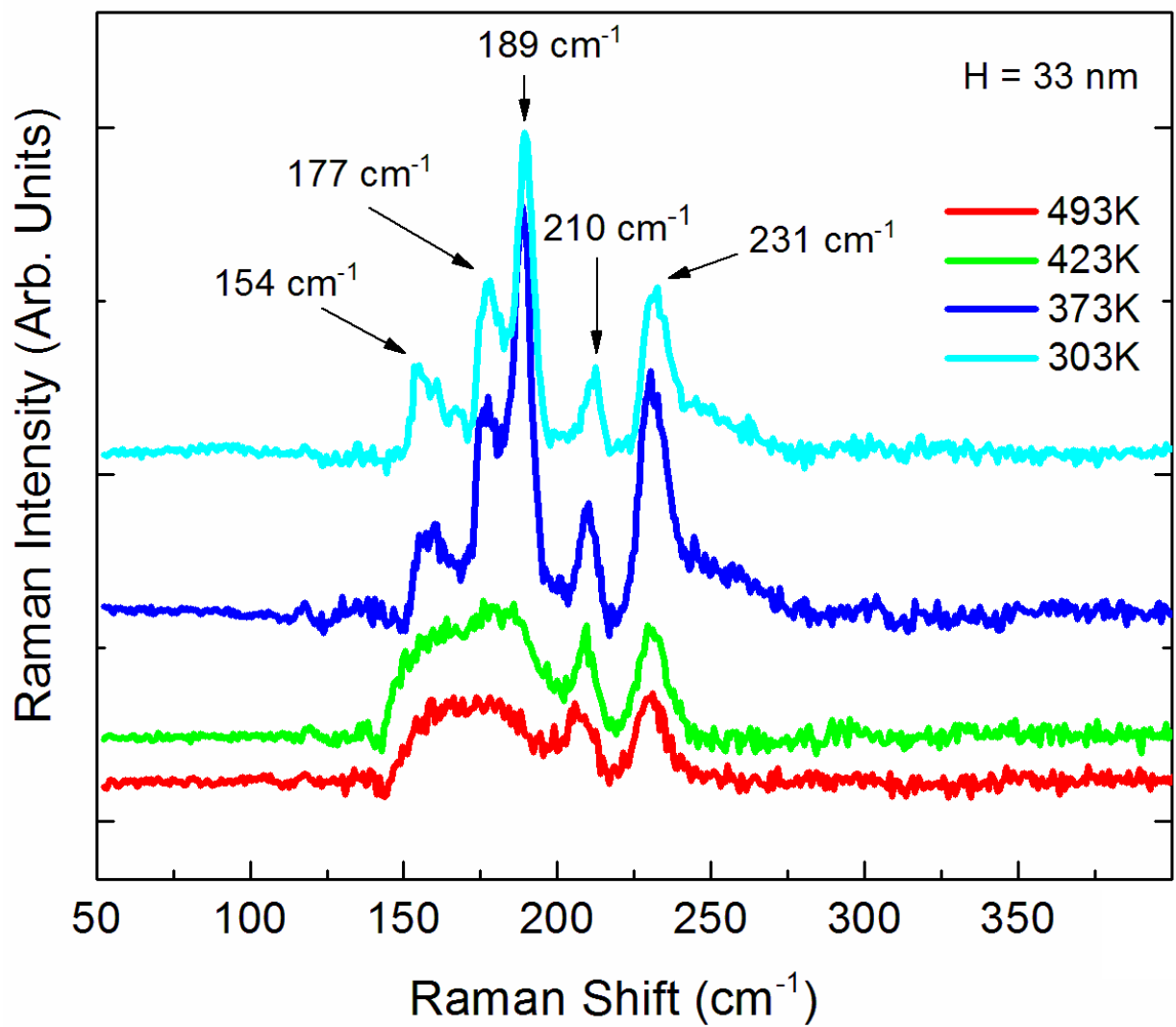

Figure 7

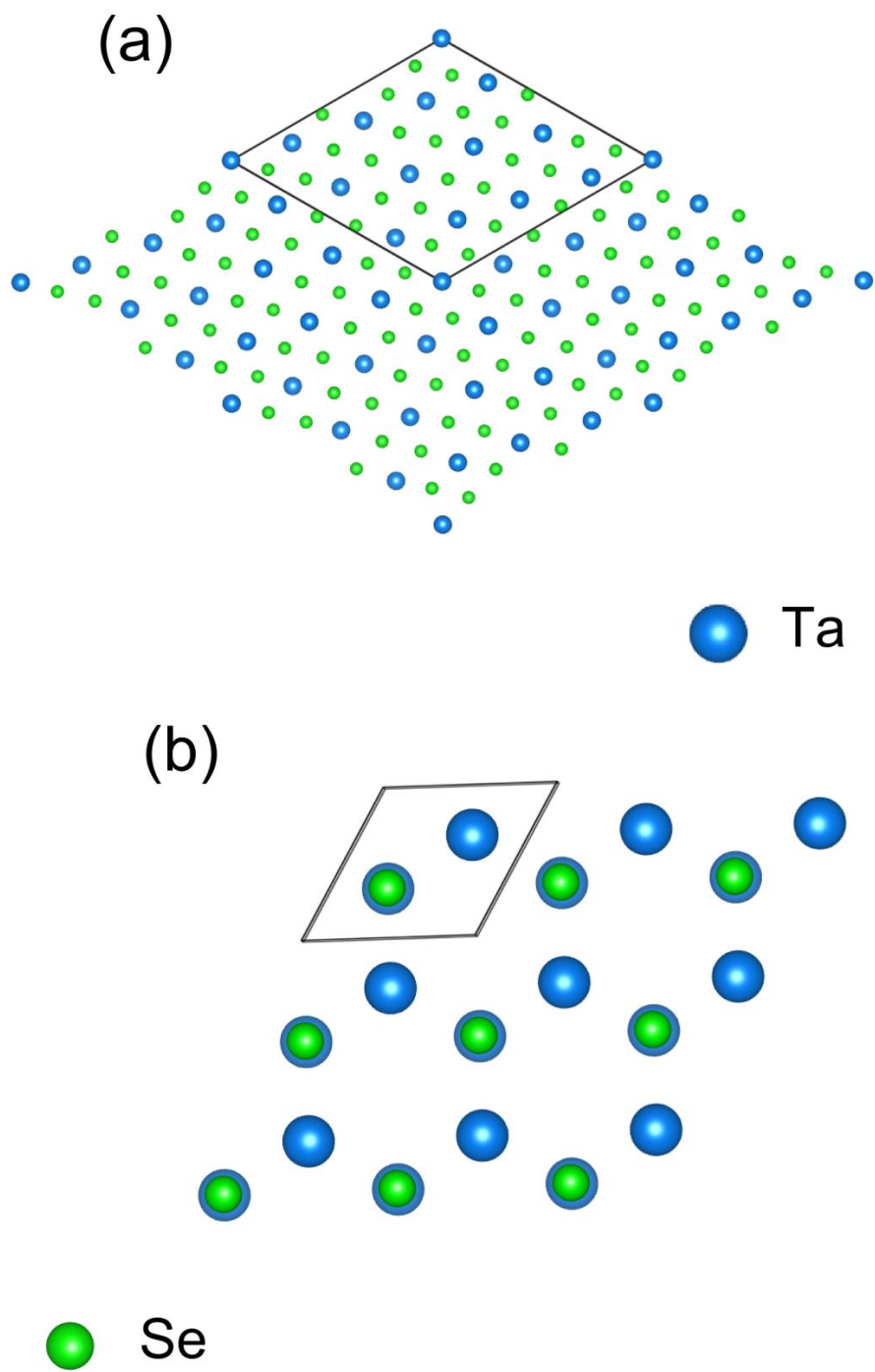

Figure 8

R. Samnakay, D. Wickramaratne, T. R. Pope, R. K. Lake, T. T. Salguero and A.A. Balandin (2014)

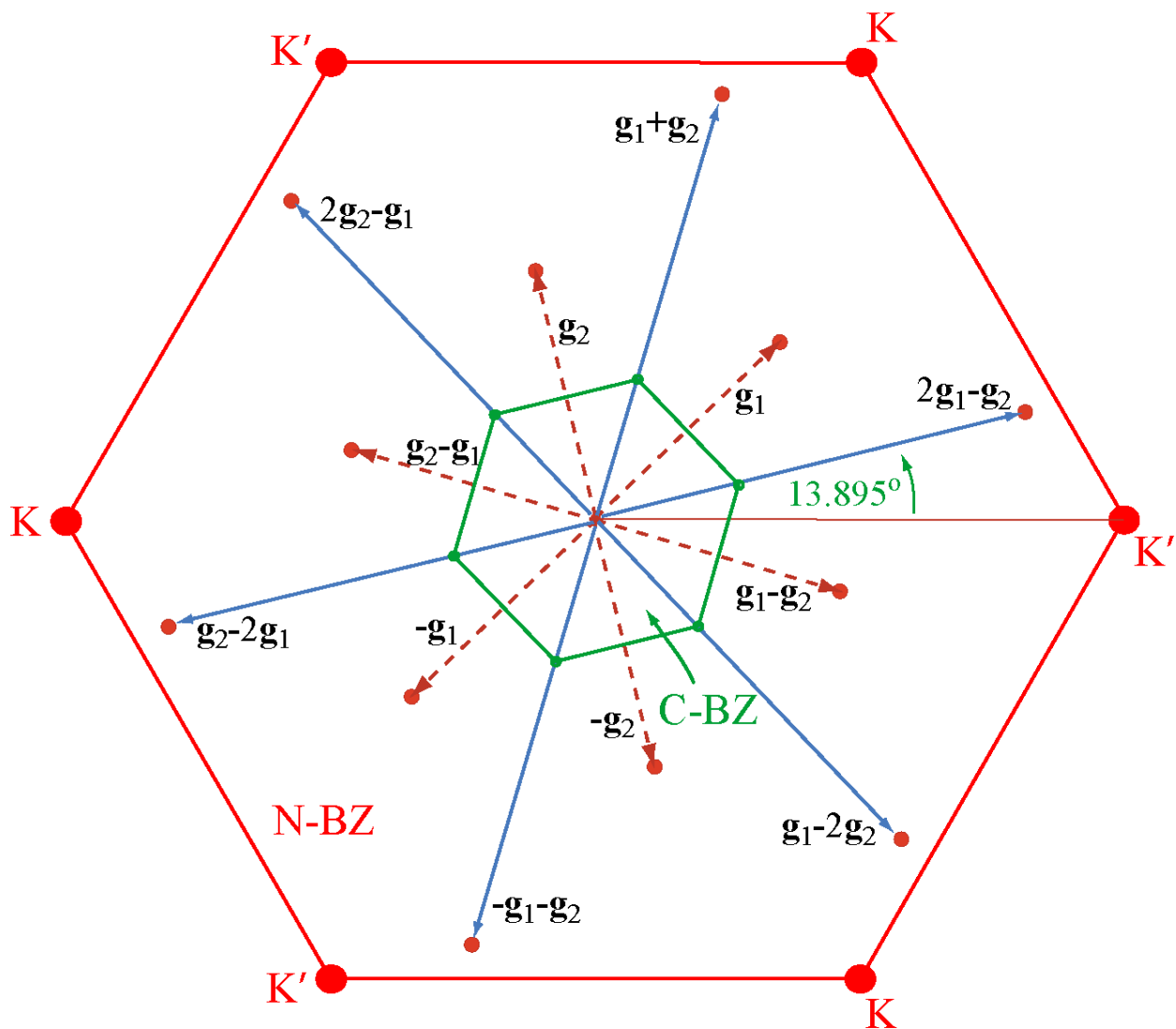

Figure 9